\documentclass[aps,twocolumn,floats,prd,nofootinbib,superscriptaddress,10pt]{revtex4-1}
\usepackage[utf8]{inputenc}
\usepackage{bm}
\usepackage{amsmath}
\usepackage{comment}
\usepackage{color}

\usepackage{natbib, eucal, cancel, float, graphicx,amsmath,amsfonts,amssymb}
\usepackage[utf8]{inputenc}

\begin{document}
\title{Direct millicharged dark matter cannot explain EDGES}

\author{Cyril Creque-Sarbinowski}
\email{creque@jhu.edu}
\affiliation{Department of Physics and Astronomy, Johns Hopkins
     University, 3400 N.\ Charles St., Baltimore, MD 21218, USA}
\author{Lingyuan Ji}
\email{lingyuan.ji@jhu.edu}
\affiliation{Department of Physics and Astronomy, Johns Hopkins
     University, 3400 N.\ Charles St., Baltimore, MD 21218, USA}
\author{Ely D. Kovetz}
\email{kovetz@bgu.ac.il}
\affiliation{Department of Physics and Astronomy, Johns Hopkins
     University, 3400 N.\ Charles St., Baltimore, MD 21218, USA}
\affiliation{Department of Physics, Ben-Gurion University of the Negev, Be'er Sheva 84105, Israel}
\author{Marc Kamionkowski}
\email{kamion@jhu.edu}
\affiliation{Department of Physics and Astronomy, Johns Hopkins
     University, 3400 N.\ Charles St., Baltimore, MD 21218, USA}

\begin{abstract}

Heat transfer between baryons and millicharged dark matter has been invoked as a possible explanation for the anomalous 21-cm absorption signal seen by EDGES.  Prior work has shown that the solution requires that millicharged particles make up only a fraction $(m_\chi/{\rm MeV})\, 0.0115\% \lesssim f \lesssim 0.4\%$ of the dark matter and that their mass $m_\chi$ and charge $q_\chi$ have values $0.1\lesssim (m_\chi/{\rm MeV}) \lesssim 10$ and $10^{-6} \lesssim (q_\chi/e) \lesssim 10^{-4}$.  Here we show that such particles come into chemical equilibrium before recombination, and so are subject to a constraint on the effective number $N_{\rm eff}$ of relativistic degrees of freedom, which we update using Planck 2018 data.  We moreover determine the precise relic abundance $f$ that results for a given mass $m_\chi$ and charge $q_\chi$ and incorporate this abundance into the constraints on the millicharged-dark-matter solution to EDGES.  With these two results, the solution is ruled out if the relic abundance is set by freeze-out. 

\end{abstract}

\maketitle

\section{Introduction}

The global 21-cm signal centered at 78 MHz was reported by the
Experiment to Detect the Global Epoch of Reionization Signature
(EDGES)~\cite{1708.05817} to be more than twice as deep
than allowed by the standard cosmological model.  This anomaly has been explained in terms of heat transfer between baryons and an interacting component of dark matter (DM) \cite{1803.06698}, as anticipated in Refs.~\cite{1408.2571,1509.00029}.  This explanation requires, though, that such an interaction increase in strength at lower baryon-DM relative velocities to evade constraints from the cosmic microwave background (CMB) \cite{1311.2937, 1802.06788, 1803.09734, 1808.00001}.  Currently, the only viable particle-physics models are those in which the interacting dark-matter component has a millicharge \cite{hep-ph/0307284, gr-qc/0411113, hep-ph/0611184, 0712.0607, 1408.3588}.

Millicharged dark matter is constrained by accelerator experiments \cite{hep-ex/9804008}, big-bang nucleosynthesis (BBN) \cite{hep-ph/0001179,1303.6270}, stellar cooling \cite{1311.2600}, and SN1987A \cite{1803.00993}.
Refs.~\cite{1802.10094,PhysRev.D98.103005, 1803.02804,1805.11616} explored the implications of these constraints for EDGES, concluding that millicharged dark matter can explain EDGES if only a small component of the dark matter interacts with baryons.
Refs.~\cite{1808.00001,1807.11482} improved and updated the CMB constraints, carefully treated the strong-coupling regime at low DM fractions, and  identified a minimum millicharged-DM fraction required to explain the EDGES signal. As a result, the current viable millicharged-dark-matter parameter space is limited to masses $0.1 \lesssim (m_\chi/{\rm MeV}) \lesssim 10$, charges $10^{-6}\lesssim (q_\chi/e) \lesssim 10^{-4}$, and fractions $(m_\chi/{\rm MeV}) 0.0115\% \lesssim f \lesssim 0.4\%$, with $e$ the electron charge. Moreover, the millicharged particles must obtain their charge from the Standard Model photon, a scenario we call direct millicharged dark matter.

In this paper, we first determine the millicharged-DM abundance by thermal freeze-out for a given mass and charge, and consider the implications for the parameter space of Ref.~\cite{1807.11482}.  We moreover verify the  chemical equilibrium assumption used in the recombination constraint to light millicharged-DM \cite{1303.6270} and update it with current Planck 2018 data \cite{1807.06209}. We find with these new results that if the millicharged-DM abundance is fixed by thermal processes, and no additional interactions (such as involving neutrinos) are present, then the millicharged-DM explanation of EDGES is ruled out.

This paper is organized as follows:  In Section~\ref{sec:therm}, we verify analytically and numerically the validity of the assumption that the millicharged particles are in chemical equilibrium within the relevant parameter space. Then, in Section~\ref{sec:relic} we relate the fraction $f$ of DM today to the mass and charge of the particle through freeze-out. Finally, in Section~\ref{sec:Neff} we reproduce the calculations done in Ref.~\cite{1303.6270} with Planck 2018 data.  We discuss and conclude these results in Section~\ref{sec:conc} and Section~\ref{sec:disc}. 
\section{Thermalization}\label{sec:therm}
\begin{figure}[h!]
\includegraphics[width = 0.9\linewidth, height = 0.7\linewidth]{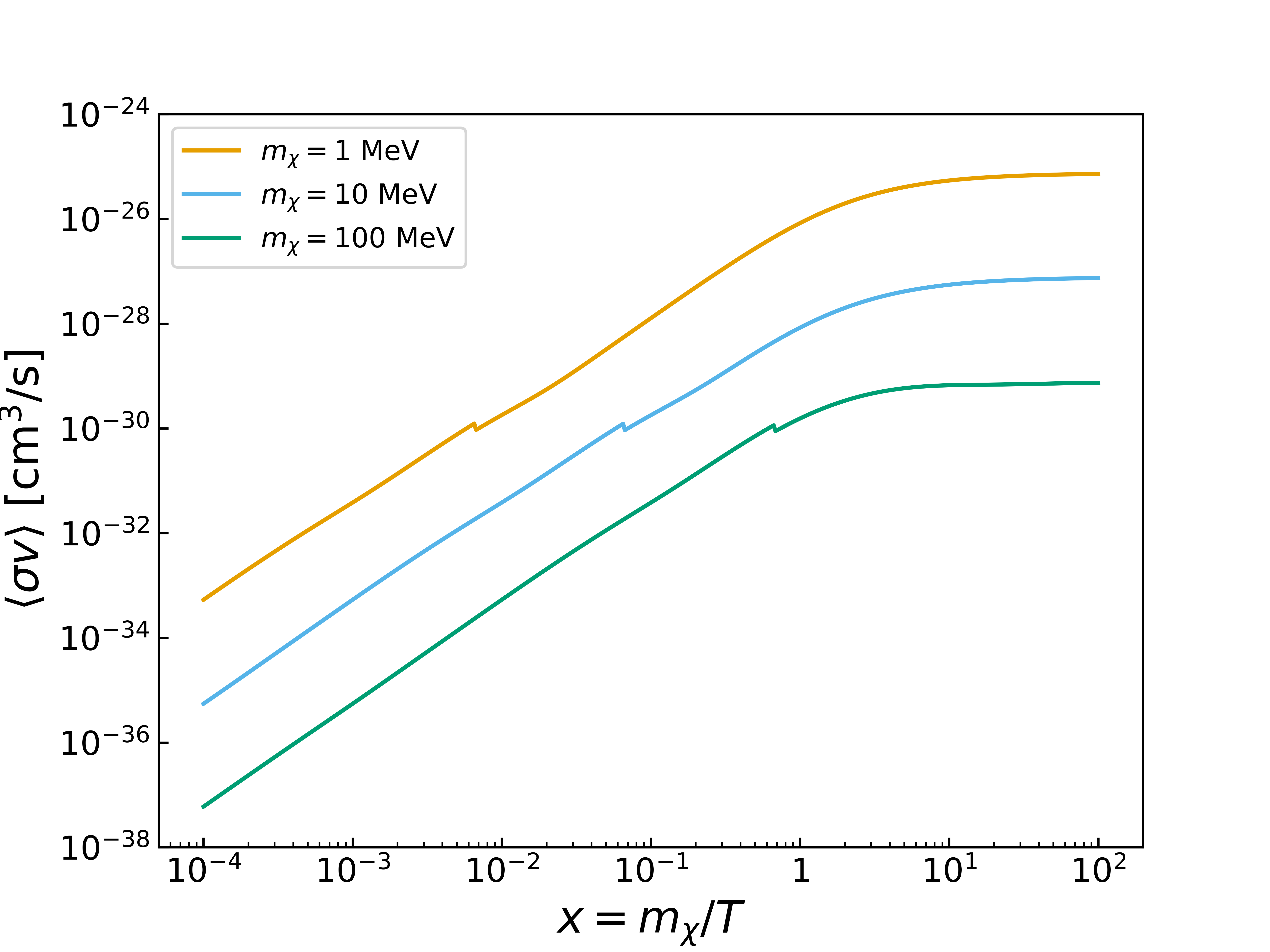}
\caption{The sum of thermal cross sections between all charged Dirac fermions in the Standard Model and a millicharged particle with mass $m_\chi \in \{1\ {\rm MeV}, 10\ {\rm MeV}, 100\ {\rm MeV}\}$ and charge $q_\chi = 10^{-6}e$. For temperatures higher than the electron mass this expression follows the expected Coulombic scaling relation $x^2 \sim T^{-2}$, while exponentially cutting off at lower temperatures. There is a period, however, below the millicharge mass, where the thermal cross section is constant, though only when the millicharge mass is larger than the electron mass. The discrete jump is due to the change in particle content at the QCD crossover.}\label{fig:thermalcs}
\end{figure}

Consider a particle with mass $m_\chi$ and electromagnetic charge $q_\chi$. For simplicity, we will take the particle to be a Dirac fermion, but discuss the scalar case in Appendix~\ref{app:CS}.  We assume that the particle initially has zero occupation at a photon temperature higher than the particle mass. However, electromagnetic interactions with charged elementary particles increase the occupancy, which can be obtained from detailed balance of the pair-production cross section $\sigma^\alpha \equiv \sigma_{\chi\bar{\chi}\rightarrow \alpha \bar{\alpha}}$.  At tree level, this cross section is given by
\begin{align}
\frac{\sigma^\alpha}{(s + 2m_\alpha^2)(s + 2m_\chi^2)} = N_c^2\frac{q_\alpha^2 q_\chi^2}{12\pi s^3}\sqrt{\frac{1 - 4(m_\alpha^2/s)}{1 - 4(m_\chi^2/s)}},
\end{align}
where $m_\alpha$ and $q_\alpha$ are the mass and charge of another charged Dirac fermion $\alpha$, $s$ the center-of-mass energy squared, and $N_c$ the number of colors (three for quarks, one for all others). We neglect photon annihilation as the cross section is two orders higher in $q_\chi$. The relevant quantity for the production of a population of millicharged particles however is the thermally averaged cross section $\langle \sigma v\rangle = \sum_\alpha \langle \sigma^\alpha v \rangle$~\cite{Gondolo:1990dk},
\begin{align}\label{eq:sigmav}
\nonumber \langle \sigma^\alpha v \rangle &= \frac{1}{8m_\chi^4 T K_2^2(m_\chi/T)}\\
&\int_{4\max\left(m_\chi, m_\alpha\right)}^\infty ds \sqrt{s}(s-4m_\chi^2)\sigma^\alpha K_1(\sqrt{s}/T),
\end{align} 
with $T$ the photon temperature, and $K_i(x)$ the modified Bessel function of order $i$. We plot Eq.~\eqref{eq:sigmav} in Fig.~\ref{fig:thermalcs} after summing over all charged Dirac fermions in the Standard Model.  

Millicharged particles are created by the annihilation of Dirac fermions and depleted by the inverse reaction.  The abundance $n_\chi$ of millicharged particles is thus governed by the Boltzmann equation
\begin{align}\label{eq:boltzmann}
\frac{dY_\chi}{dx} &= -\lambda\left[Y_\chi^2 - (Y_\chi^{\rm eq})^2\right],
\end{align}
written in terms of $Y_\chi = n_\chi/ s$, and its equilibrium counterpart $Y_\chi^{\rm eq} = (n_\chi^{\rm eq})/s$.  Here, $\lambda = s \langle \sigma v \rangle / (dx/dt)$ with $dx/dt = x s\sqrt{3 \rho}  /(M_{\rm pl}c)$ and $M_{\rm pl}$ is the reduced Planck mass.  In addition, $\rho = (\pi^2/30) g_\rho(T) T^4$ is the energy density, $s=(2\pi^2/45) g_s(T) T^3$ the entropy density, and $c=T(ds/dT) = (2\pi^2/15)g_c T^3$ the heat-capacity density \cite{1204.3622}.  We use the dimensionless inverse temperature $x = m_\chi/ T$ to track time, and the values from Ref.~\cite{1606.07494} for $g_s$, $g_\rho$, and $g_c$, where $g_i$ is the relativistic degrees of freedom for the corresponding density $i$. 

We now investigate if millicharged particles reach chemical equilibrium by determining when pair production is efficient. That is, if the number of millicharged particles $dY_\chi$ created  in some fraction of a time $dx/x$ is greater than or equal to $Y_\chi^{\rm eq}$, the chemical equilibrium number, then this process is efficient and chemical equilibrium is reached instantaneously. Otherwise, chemical equilibrium is not reached. Since the last particle in the Standard Model to go non-relativistic is the electron, the latest time (or lowest photon temperature) that chemical equilibrium could be obtained is at $T_{\rm min} \approx \max(m_\chi, m_e)$. At smaller temperatures, either the equilibrium abundance or the thermal cross section exponentially cuts off and production of millicharged particles is suppressed. 

 For millicharged particle masses $0.1\lesssim (m_\chi/{\rm MeV}) \lesssim 100$, the minimum temperature is approximately the millicharged particle mass, $T_{\rm min} \approx m_\chi$, and the only relevant thermal cross section is with electrons, $\langle \sigma v \rangle \approx \langle \sigma^e v\rangle$. Therefore, as long as reheating, or any other particle-production mechanism, produces a thermal bath containing at least photons and electrons at temperature $T \geq T_{\rm min}$, it is possible to attain chemical equilibrium with this bath. Moreover, at such a temperature the thermal cross section simplifies as $\langle \sigma v \rangle \approx \langle \sigma^e v\rangle \approx q_\chi^2 e^2/(16\pi^2 T^2)$. We then evaluate and rearrange the aforementioned condition $(dY_\chi/d\log x)/Y_\chi^{\rm eq} \gtrsim 1$. Assuming we have not reached chemical equilibrium, $Y_\chi^2 \ll \left(Y_\chi^{\rm eq}\right)^2$, the temperature of equilibration has an upper bound,  
\begin{equation}\label{eq:thermtemp}
T_{\rm eq} \lesssim 100\ {\rm GeV}\left(\frac{q_\chi}{10^{-6} e}\right)^2\left(\frac{g_c(T_{\rm eq})}{g_s(T_{\rm eq})}\right)\left(\frac{10}{g_\rho(T_{\rm eq})}\right)^{\frac{1}{2}}.
\end{equation}
Evaluating Eq.~\eqref{eq:thermtemp} at the minimum temperature $T_{\rm eq} = T_{\rm min}$ allows us to characterize equilibration only in terms of the millicharged particle's mass $m_\chi$ and charge $q_\chi$,
\begin{equation}\label{eq:thermalize}
\frac{q_\chi}{e} \gtrsim 10^{-8.5}\left(\frac{g_\rho(m_\chi)}{10}\right)^{\frac{1}{4}}\left(\frac{g_s(m_\chi)}{g_c(m_\chi)}\right)^{\frac{1}{2}}\left(\frac{m_\chi}{1\ {\rm MeV}}\right)^{\frac{1}{2}}.
\end{equation}  
It follows that an initial millicharged abundance of zero will still reach chemical equilibrium within the allowed region of masses and charges. In order to verify the above conditions, we now perform a numerical check using Eq.~\eqref{eq:boltzmann}, which involves all relevant Standard Model particles. 

\begin{figure}
\includegraphics[width = \linewidth, height = 0.7\linewidth]{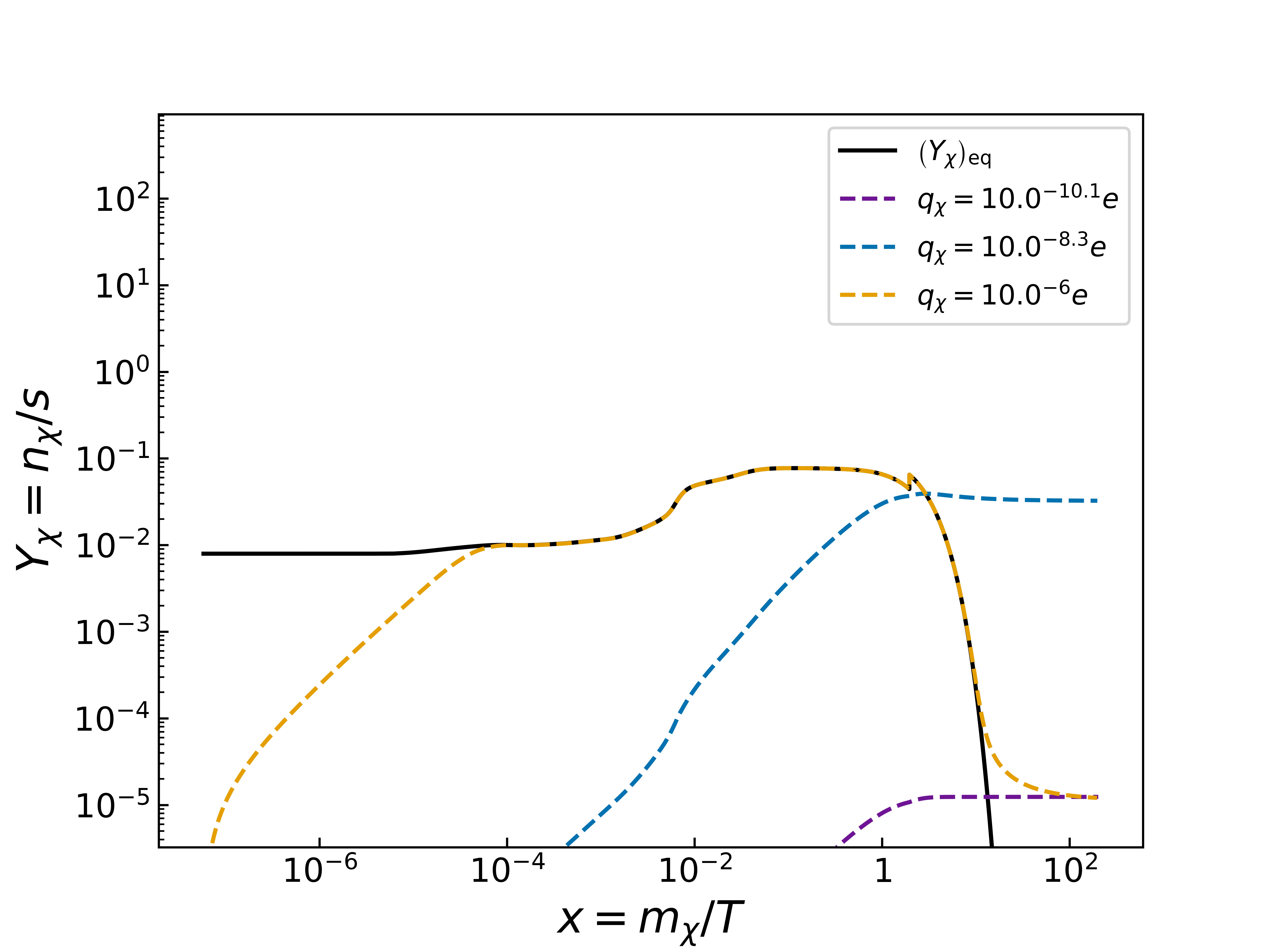}
\caption{The abundance of a $m_\chi = 1$ MeV millicharged particle with charge $q_\chi \in \{10^{-6}e, 10^{-8.3}, 10^{-10.1}e\}$, evolved with Eq.~\eqref{eq:boltzmann}. For the $q_\chi = 10^{-6}e$ case, the particle thermalizes at around $100$ GeV. Otherwise, it never reaches chemical equilibrium, represented by $Y_\chi^{\rm eq}$. For the $q_\chi = 10^{-10.1}e$ scenario the abundance is set by freeze-{\it in}, and can achieve the same relic abundance as freeze-{\it out} with $q_\chi = 10^{-6}e$. However such charges are too small to produce the EDGES signal.}\label{fig:thermalize}
\end{figure}

First, we check that the equilibration time specified by Eq.~\eqref{eq:thermtemp} is correct. Then, we check that the boundary between equilibration and non-equilibration in Eq.~\eqref{eq:thermalize} is correct. Finally, we double check that in our region of parameter space chemical equilibrium is achieved. However, we only check that this equilibration occurs for a particle with the smallest permissible charge, as all other points have higher production efficiencies at $T_{\rm min}$. We demonstrate all three checks in Fig.~\ref{fig:thermalize} and find they all clear. Although it is not plotted, the same conclusion holds for millicharged particles that are complex scalars. 

We conclude that in the region of interest, millicharged particles reach chemical equilibrium and undergo freeze-out. 
\begin{figure}
\includegraphics[width = \linewidth]{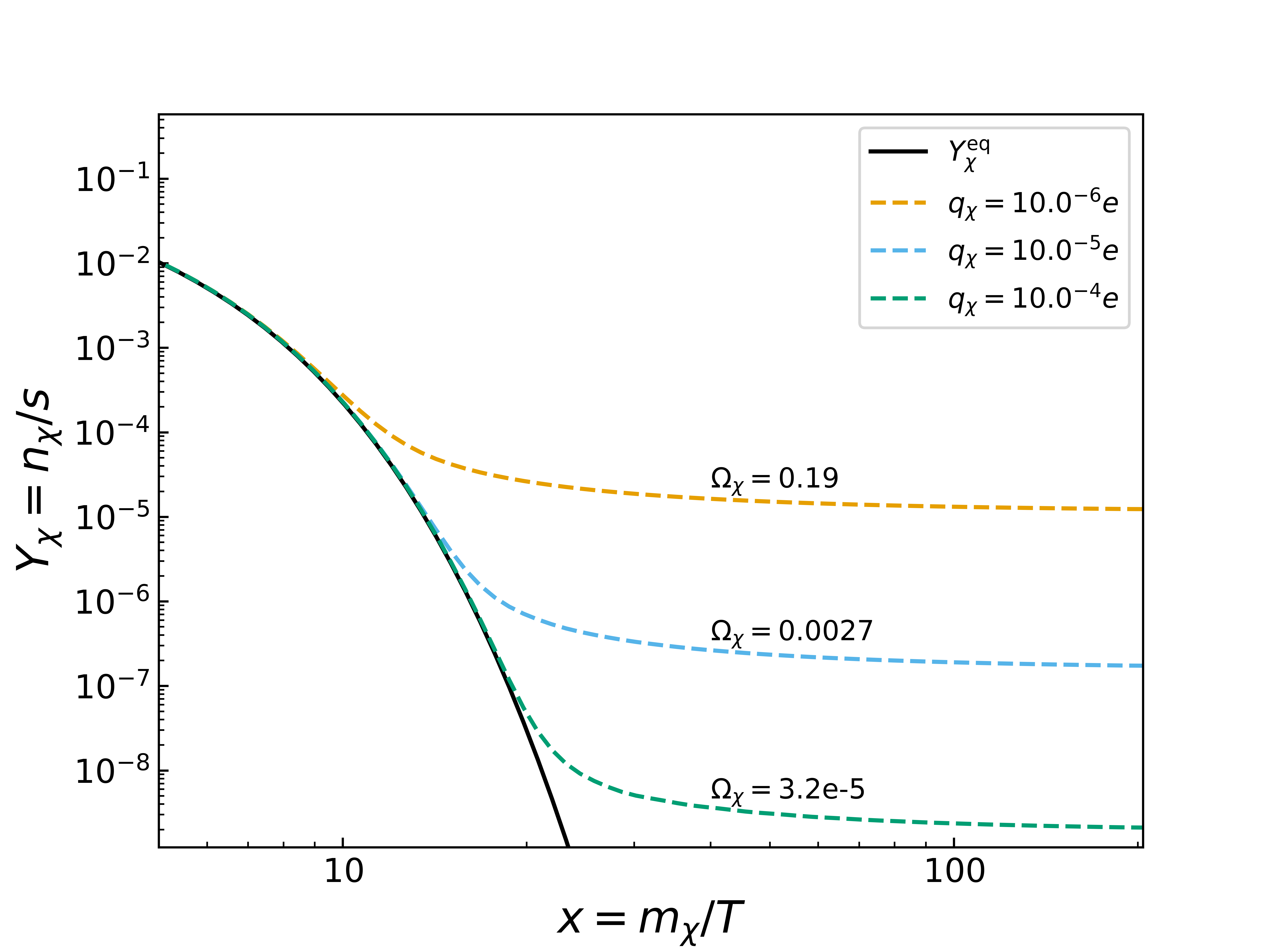}
\caption{The numerically calculated freeze-out abundances $\Omega_\chi$ of a Dirac fermion with mass $m_\chi = 1\ {\rm MeV}$ and charge $q_\chi \in \{10^{-6}e, 10^{-5}e, 10^{-4}e\}$. Its equilibrium abundance $Y_\chi^{\rm eq}$ is plotted for reference.}\label{fig:finalabundance}
\end{figure}
\section{Relic Abundance}\label{sec:relic}
If no depletion branches exist, the number of millicharged particles at freeze-out uniquely determines the amount today. More specifically, as the coupling with the thermal bath increases, the number of millicharged particles after freeze-out decreases, see Fig.~\ref{fig:finalabundance}, as the particles are in chemical equilibrium for a longer period of time. Neglecting the equilibrium term at freeze-out due to its exponential decrease, and treating the thermal cross section to be constant, we integrate Eq.~\eqref{eq:boltzmann} from the time $x_f$ of freeze-out until today (which we take to be $x = \infty$). Note that the number of particles at the onset of freeze-out is much larger than the sum total today, see Fig.~\ref{fig:thermalize}, and so its inverse can be neglected post-integration. The present photon temperature $T_{\rm cmb}$ is much smaller than the millicharged particle mass today, and so the population of millicharged particles is non-relativistic. Therefore, we convert the relic number of millicharged particles $Y_\chi(x = \infty)$ into an energy density by multiplying both by its mass and the current entropy density. This multiplication leads us to express the energy density today in units of the critical energy density as 
\begin{equation}\label{eq:relicabundance}
\Omega_\chi = \frac{\pi}{9}\frac{x_f}{\langle \sigma v \rangle}\left(\frac{g_\rho(m_\chi)}{10}\right)^{1/2}\frac{g_s(T_{\rm cmb})}{g_{c}(m_\chi)}\frac{T_{\rm cmb}^3}{M_{\rm pl}^3 H_0^2},
\end{equation}
with $H_0$ the Hubble constant today. This equation holds for both Dirac fermions and complex scalars. Ref.~\cite{1807.11482} reported this abundance must compromise a fraction between $(m_\chi/{\rm MeV})\ 0.0115\%\lesssim f \lesssim 0.4 \%$ of the entire DM content in order to both explain the EDGES 21-cm signal and evade the CMB constraints on the model. 

\begin{figure}[H]
\includegraphics[width = \linewidth]{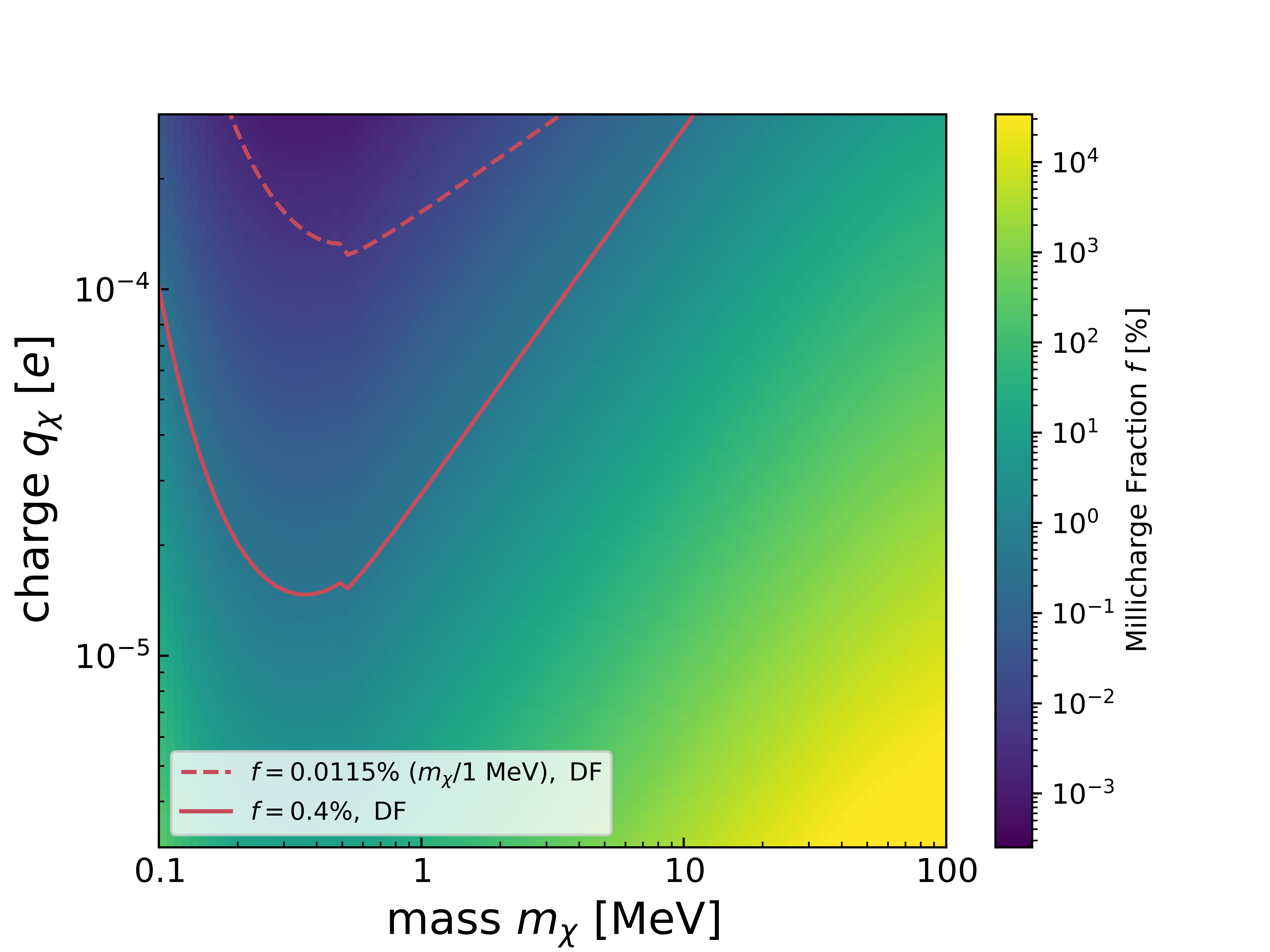}
\caption{The dark-matter fraction $f$ of millicharged particles with mass $m_\chi$ and charge $q_\chi$ for Dirac fermions (DF). At fixed mass, the abundance decreases with increasing charge. At fixed charge the abundance minimizes at the electron mass due to a peak in the cross section with electrons, and increases on either side otherwise. The sudden jump at around the electron mass is due to the assumed discrete change in temperature after $e^{\pm}$ annihilation. }\label{fig:relicabundance}
\end{figure}

In Fig.~\ref{fig:relicabundance} we plot the millicharged fraction $f = \Omega_\chi/ \Omega_c$, with $\Omega_c$ the cold dark matter energy density in units of the critical density. Overlaid on top are the aforementioned EDGES compatibility requirements. Since Eq.~\eqref{eq:relicabundance} is linear in the onset of freeze-out we do not worry about the exact timing of onset and take $x_f = 10$.

\vspace{0.2in}

Even after imposing the compatibility requirements, there  remains a nonzero amount of parameter space that is still viable to explain the EDGES signal. In order to constrain this remaining amount, we calculate the effect of an additional particle on the effective number of relativistic degrees of freedom during recombination. To this end, we use our earlier result that this region is in chemical equilibrium. As a result, we can use the equations of Ref.~\cite{1303.6270} that detail such an effect for particles in chemical equilibrium, only updating their calculations using Planck 2018 parameters.
\section{$N_{\rm eff}$ Bound}\label{sec:Neff}
The addition of a new particle whose non-gravitational interaction is solely electromagnetic and decoupling period is during or after neutrino decoupling $T_{\rm D}$ further enhances the photon temperature relative to the neutrino temperature $T_\nu$ due to entropy conservation. As a result, the measured effective number $N_{\rm eff}$ of relativistic degrees of freedom at recombination is shifted downward. In the context of $n$ particles with masses and degrees of freedom $\{m_i, (g_\rho)_i\}, i \in \{1,...,n\}$ and instantaneous neutrino decoupling, we use Eq.~(10) of Ref.~\cite{1303.6270} to express $N_{\rm eff}$ as 
\begin{align}\label{eq:Neff}
\nonumber N_{\rm eff} &= N_\nu\left[1 + \frac{7}{22}\sum_{i = 1}^n\frac{(g_\rho)_i}{2}F\left(\frac{m_i}{T_{\rm D}}\right)\right]^{-4/3},\\
F(x) &\equiv \frac{30}{7\pi^4}\int_x^\infty dy\frac{(4y^2 - x^2)\sqrt{y^2 - x^2}}{e^y \pm 1},
\end{align} 
with $N_\nu$ the number of relativistic neutrinos at recombination, and the plus/minus for fermionic/bosonic statistics. Realistically, the additional particle not only imposes changes in $N_{\rm eff}$ but also in the helium mass fraction $Y_{\rm P}$, due to interactions during big-bang nucleosynthesis (BBN). Although we do not calculate this mass fraction here, a proper treatment of constraining $N_{\rm eff}$ requires us to use the joint analysis on both $N_{\rm eff}$ and $Y_{\rm P}$ from Planck 2018, which allowed both variables to vary freely. In this analysis they inferred a conservative $95\%$ confidence level constraint on the effective number of relativistic degrees of freedom $N_{\rm eff} = 2.97^{+0.58}_{-0.54}$~\cite{1807.06209}. Since the effect of an additional particle is to lower $N_{\rm eff}$, we consider the Planck 2018 lower bound when inferring the CMB limit.

\begin{figure}
\includegraphics[width = \linewidth]{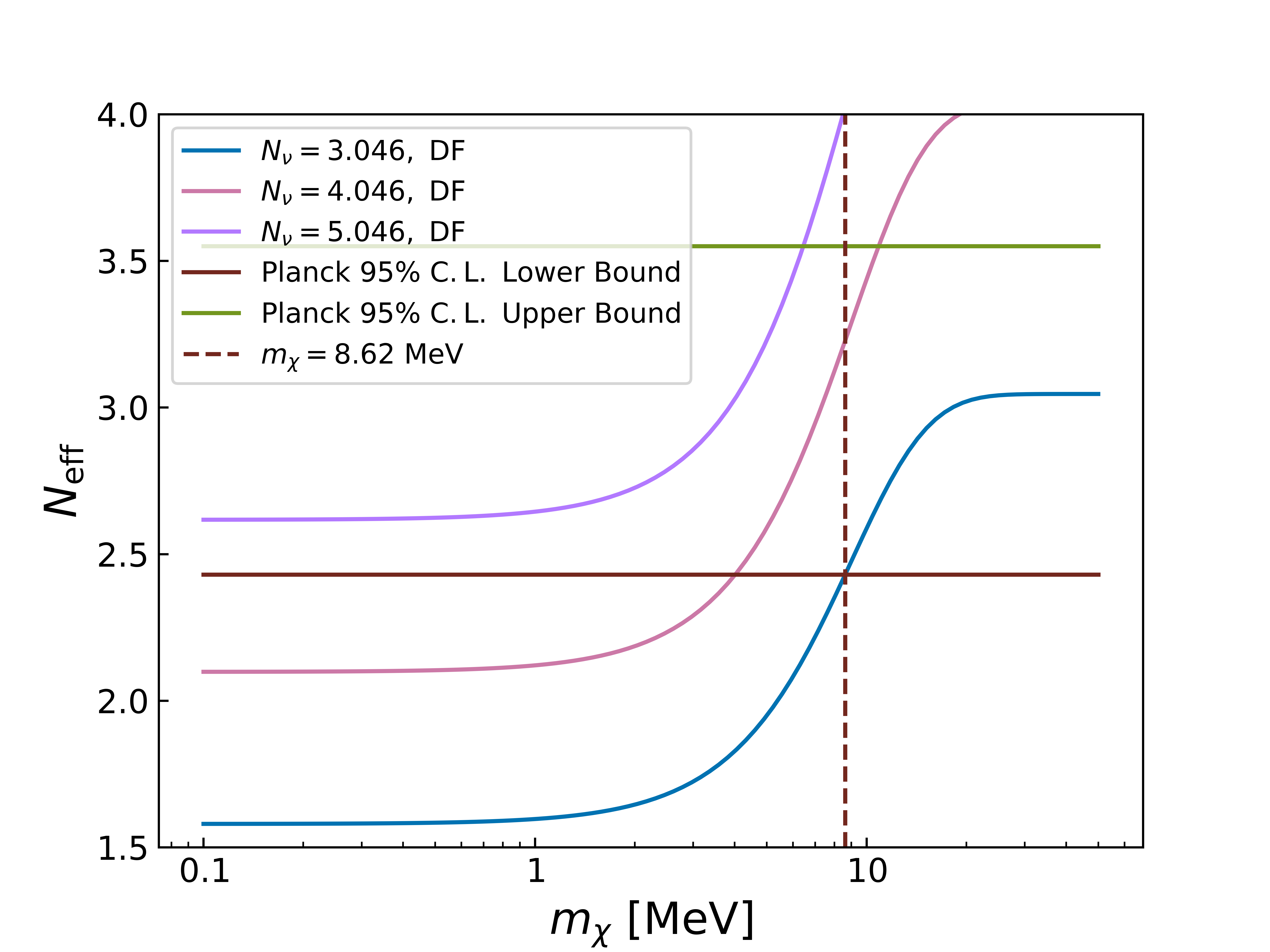}
\caption{The effective number $N_{\rm eff}$ of relativistic degrees of freedom as a function of the millicharged particle mass $m_\chi$, assuming $N_\nu$ relativistic neutrinos at recombination for a Dirac fermion (DF). The solid reddish brown line is the 95\% confidence level lower bound from Planck 2018. The dashed counterpart is the resulting lower bound on the millicharged particle mass for $N_\nu = 3.046$.}\label{fig:Neff}
\end{figure}

In order to derive constraints on our millicharged particle ($(g_\rho)_1 = 2\ {\rm or}\ 4)$, we plot in Fig.~\ref{fig:Neff} both this lower bound as well as Eq.~\eqref{eq:Neff},  taking $T_{\rm D} = 2.3$ MeV and $N_\nu = 3.046$. In addition, we show via the same plot that it is possible to evade the Planck 2018 lower bound constraint if the Universe has two extra neutrinos at the time of recombination for a Dirac fermion (DF). Finally, for reference we show the Planck 2018 upper bound on $N_{\rm eff}$. 

\begin{figure}[H]
\includegraphics[width = \linewidth]{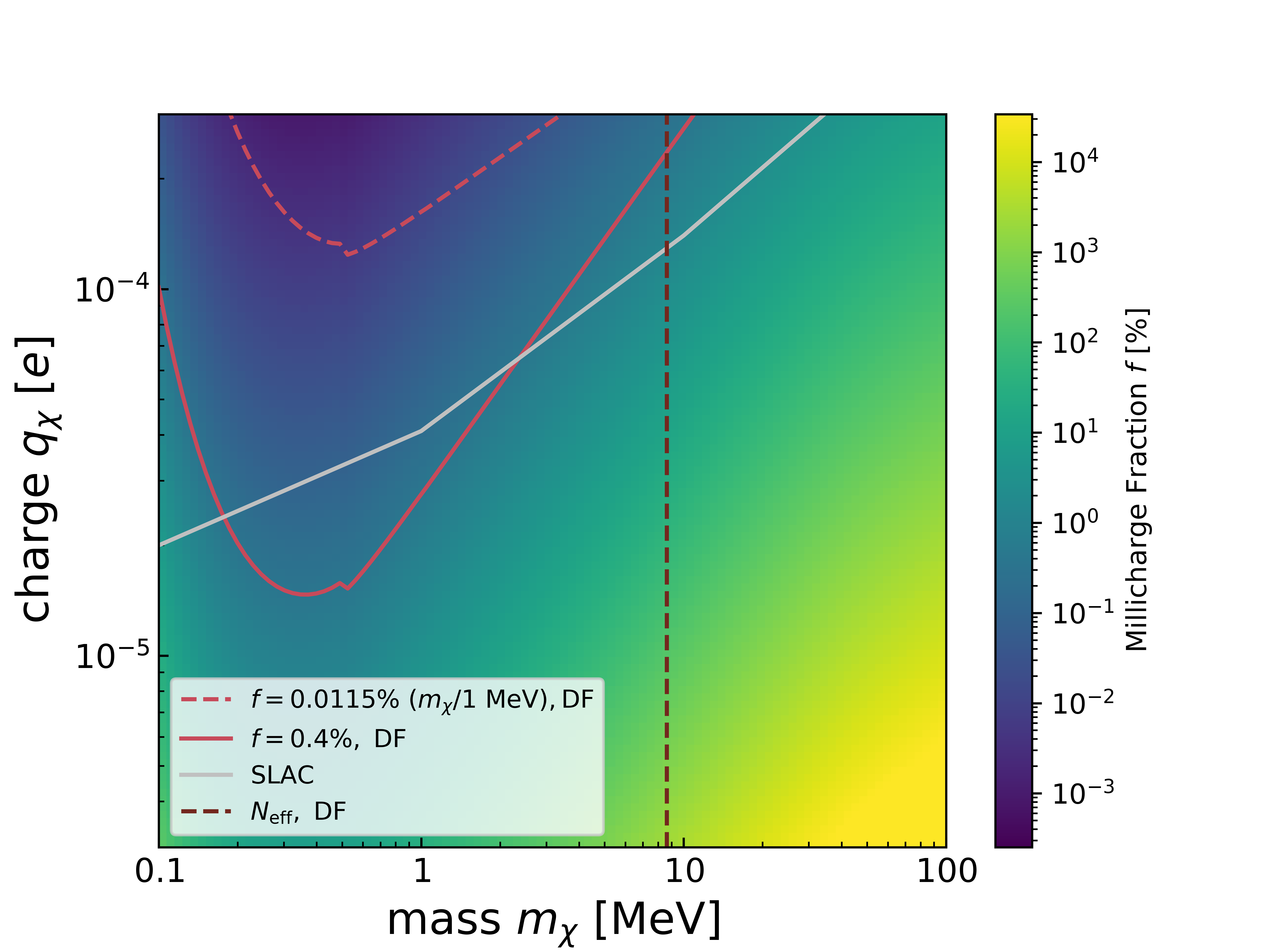}
\caption{The dark matter fraction $f$ of millicharged particles with mass $m_\chi$ and charge $q_\chi$ for Dirac fermions (DF). The region between the dashed and solid red line is the range of relic abundances that are compatible with CMB and EDGES constraints. The region above the solid grey line is ruled out due to SLAC measurements. Finally, the region to the left of the reddish brown vertical line is ruled out due to $N_{\rm eff}$ constraints. The viable regions between the SLAC and relic abundance regions have nonzero overlap, but this overlapped region does not intersect with the region permitted by $N_{\rm eff}$.}\label{fig:constraints}
\end{figure}

Since any value of $N_{\rm eff}$ below the Planck value is ruled out, we impose a lower bound on the millicharged particle mass at $m_\chi = 8.62\ {\rm MeV}$. We show this bound, along with the most recent upper bound on the charge of the millicharged particle from SLAC \cite{hep-ex/9804008}, in Fig.~\ref{fig:constraints}. Combined with our prior relic abundance constraints, and those of Ref.~\cite{1807.11482}, the millicharged particle is completely ruled out.
\section{Discussion}
\label{sec:disc}
There are a few caveats when applying our bounds. First, if there exist at least two (but no more than three) extra neutrinos, it is possible to evade the aforementioned $N_{\rm eff}$ constraints for Dirac fermions. 

One way to generate these extra degrees of freedom is by considering that the millicharged particle comes along with a kinetically mixed massless hidden photon. Ref.~\cite{PhysRev.D98.103005} already claimed to rule out this model, however this was only for large dark couplings $g'$ between the hidden photon and millicharged particle. This constraint~\cite{1311.2600} arises because large dark couplings overshoot the value of $N_{\rm eff}$ during BBN. If the dark coupling was identically zero, however, only the millicharged particle would be thermalized and the value of $N_{\rm eff}$ would be undershot, see Fig.~\ref{fig:Neff}. Thus, we ask if the increase of $N_{\rm eff}$ by the hidden photon can be balanced by a corresponding decrease from the millicharged particle. 

However, it is expected that the kinematic mixing parameter $\chi = q_\chi/g'$ is less than $10^{-2}$. In addition, only charges $q_\chi \gtrsim 10^{-6}e$ are viable to explain EDGES. Taken together, this leads us to conclude that $g'\gtrsim 10^{-4}e$. For such values, Section 4.2 of Ref.~\cite{1311.2600} shows that their bounds on $N_{\rm eff}$, which exclude the parameter space of interest, still apply. 

Secondly, atomic dark matter, a scenario where a residual free dark electron fraction follows a dark recombination, may still be viable \cite{0909.0753, 1201.4858, 1209.5752}. The remaining ionized DM could provide the fractional millicharged DM component that is required to explain the EDGES signal\footnote{Curiously, the ionized fraction of the baryonic gas around cosmic dawn is of order $10^{-4}$, similar to the values required to explain EDGES via millicharged DM (see Ref. [22]).}.

Moreover, an additional interaction could exist between neutrinos and the millicharged particle that is efficient during millicharged particle annihilation. In this case, the heat from annihilation would not only go to photons, but also to neutrinos, producing no change in $N_{\rm eff}$. However, such a case would induce a coupling between charged particles and neutrinos through loop effects. As a result, it faces strong constraints from bounds on the electric charge of the neutrino \cite{PhysRev.D98.030001}.

It could also be the case that millicharged particles receive an electronic charge not only through kinetic mixing, but also through the Standard Model photon by some higher-energy physics. If the Standard Model photon charge is larger than the charge generated through kinetic mixing, so that $q_\chi \neq g'\chi$, then the dark coupling $g'$ evades the $q_\chi \gtrsim 10^{-4}e$ requirement. Thus, the $N_{\rm eff}$ change due to millicharged particles could be offset by this hidden photon. 

Finally, millicharged particles could have their abundance set not thermally, but through reheating that occurs at a temperature above BBN but below the millicharged mass \cite{astro-ph/0002127, hep-ph/0103272, astro-ph/0505395}. As a result, not only are annihilations severly supressed so that any change in $N_{\rm eff}$ is small, but also the relic abundance formula we wrote down does not hold. 
\section{Conclusion}
\label{sec:conc}
We have considered the prospects of a millicharged particle directly charged under the Standard Model photon to explain the anomalous EDGES 21-cm signal. Specifically, we  had three tasks in mind. First, we wished to verify that millicharged particles that have their abundances set thermally reach chemical equilibrium regardless of initial conditions. If so, we then wanted to calculate the millicharged-DM abundance set by thermal freeze-out for a given mass and charge in order to consider its implications for the parameter space of Ref.~\cite{1807.11482}. Lastly, we sought to improve on the Planck $N_{\rm eff}$ constraint from Ref.~\cite{1303.6270} with Planck 2018 data. 

We found that regardless of the initial abundance, millicharged particles reach chemical equilibrium and then undergo freeze-out. This evolution occurs as long as there exists a thermal photon and electron bath at a temperature higher than the millicharged particle mass. Using the Boltzmann equation, we then calculated both numerically and analytically the millicharged relic abundance and found a reduced, but still viable, portion of parameter space remaining to explain the EDGES signal. 

In order to cut down on this space further, we considered the effect of entropy dumping on the effective number $N_{\rm eff}$ of relativistic degrees of freedom. We found that this number decreases due to the increase in photon temperature after the millicharged particle decouples post-neutrino decoupling. This decrease was severe enough such that the remaining amount of parameter space was completely ruled out. 

Barring the caveats mentioned in the Section~\ref{sec:disc}, we therefore conclude that a millicharged particle cannot produce the 21-cm signal observed at EDGES. 

\begin{acknowledgements}
We thank Kimberly Boddy, Julian Mu\~noz, and Vivian Poulin for useful discussions.  CCS Acknowledges the support of the Bill and Melinda Gates Foundation. This work was supported at Johns Hopkins by NASA Grant No. NNX17AK38G, NSF Grant No. 1818899, and the Simons Foundation. 
\end{acknowledgements}
\appendix
\section{Complex Scalars}\label{app:CS}
\begin{figure}
\includegraphics[width = \linewidth]{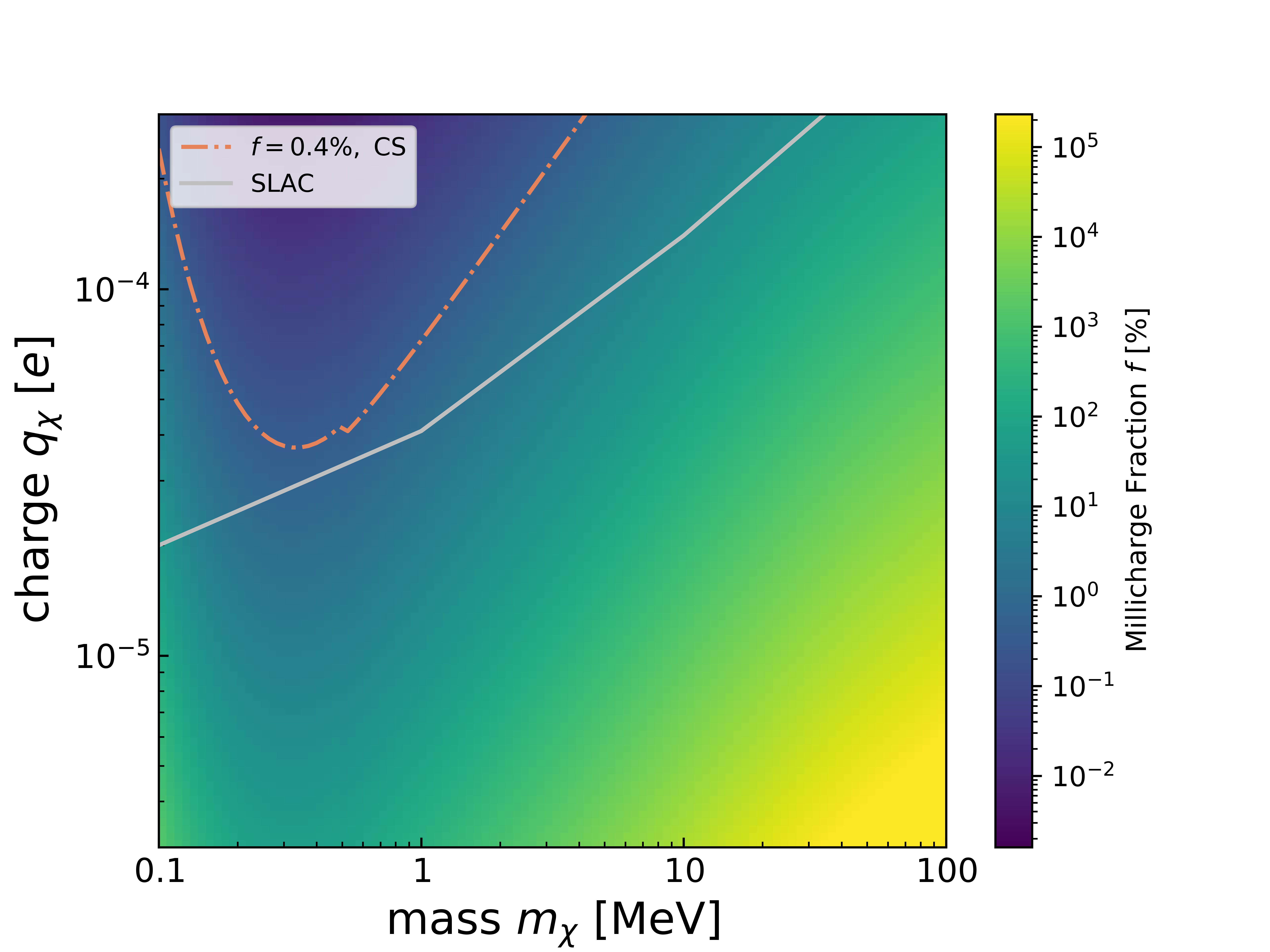}
\caption{The dark-matter fraction $f$ of millicharged particles with mass $m_\chi$ and charge $q_\chi$ for complex scalars (CS).}
\label{fig:relicabunCS}
\end{figure}

In this Appendix, we consider the millicharged particle creating the anomalous EDGES 21-cm signal to be a complex scalar (CS). Its pair-production cross section $\sigma^\alpha_{\rm CS} \equiv \sigma^{\rm CS}_{\chi\bar{\chi}\rightarrow\alpha\bar{\alpha}}$ with a Dirac fermion $\alpha$ is then 
\begin{align}\label{eq:CScross}
\frac{\sigma^\alpha_{\rm CS}}{(s + 2m_\alpha^2)(s - 4m_\chi^2)} = N_c^2\frac{q_\alpha^2 q_\chi^2}{48\pi s^3}\sqrt{\frac{1 - 4(m_\alpha^2/s)}{1 - 4(m_\chi^2/s)}}.
\end{align}
As we remarked in Section~\ref{sec:therm}, the complex scalar also reaches thermalization with the thermal bath of Standard Model particles. Thus, we can use Eq.~\eqref{eq:relicabundance} in conjuction with Eq.~\eqref{eq:CScross} to calculate the freeze-out relic abundance of these particles today, plotted in Fig.~\ref{fig:relicabunCS}.
We find that the largest allowable relic abundance by the CMB, $f \simeq 0.4$~\cite{1808.00001}, corresponds to masses $m_\chi$ and charges $q_\chi$ already ruled out by SLAC. Therefore, we conclude that a complex scalar millicharged particle cannot create the anomalous EDGES 21-cm signal.

\end{document}